**New candidates for active asteroids: main-belt (145) Adeona, (704) Interamnia, (779) Nina, (1474) Beira, and near-Earth (162173) Ryugu**

Vladimir V. Busarev[a, b,] *, Andrei B. Makalkin[c], Faith Vilas[d], Sergey I. Barabanov[b],

Marina P. Scherbina[a]

[a]*Lomonosov Moscow State University, Sternberg Astronomical Institute (SAI MSU), University Av., 13, Moscow, 119992, Russian Federation (RF)*
[b]*Institute of Astronomy, Russian Academy of Sciences (IA RAS), Pyatnitskaya St. 48, Moscow, 109017, RF*
[c]*Schmidt Institute of Physics of the Earth, Russian Academy of Sciences, Bolshaya Gruzinskaya str., 10-1, Moscow, 123242, RF*
[d]*Planetary Science Institute, 1700 E. Fort Lowell Rd., Suite 106, Tucson, Arizona, 85719, USA*

[*] Corresponding author.

E-mail address: busarev@sai.msu.ru (V.V. Busarev)

**Abstract**

For the first time, spectral signs of subtle coma activity were observed for four main-belt primitive asteroids (145) Adeona, (704) Interamnia, (779) Nina, and (1474) Beira around their perihelion distances in September 2012, which were interpreted as manifestations of the sublimation of $H_2O$ ice in/under the surface matter (Busarev et al., 2015a, 2015b). We confirm the phenomenon for Nina when it approached perihelion in September 2016. At the same time, based on results of spectral observations of near-Earth asteroid (162173) Ryugu (Vilas, 2008) being a target of Japan's *Hayabusa 2* space mission, we suspected a periodic similar transient activity on the Cg-type asteroid. However, unlike the main-belt primitive asteroids demonstrating sublimation of ices close to their perihelion distances, the effect on Ryugu was apparently registered near aphelion. To explain the difference, we calculated the subsolar temperature depending on heliocentric distance of the asteroids, considered qualitative models of internal structure of main-belt and near-Earth primitive asteroids including ice and performed some analytical estimations. Presumed temporal sublimation/degassing activity of Ryugu is a sign of a residual frozen core in its interior. This could be an indication of a relatively recent transition of the asteroid from the main asteroid belt to the near-Earth area.

# 1. Introduction

As follows from the mineralogy of carbonaceous chondrites (e. g., Dodd, 1981; Brearley, 2006) being along with other meteorites the oldest solids in the Solar System (e. g., Connelly et al., 2008; Amelin, Krot, 2007) and possible fragments of primitive asteroids (C-type and others), these asteroids are the least geologically changed solid celestial bodies. Thus, they could preserve fingerprints of the first physical, chemical and dynamical processes that formed all of the planets. Widespread condensation and accumulation of volatile compounds ($H_2O$, $CO_2$, CO, etc.) in the internal Solar System beyond the "snow-line" obviously was one of the key processes that influenced the formation of not only proto-Jupiter and other giant planets, but also a swarm of smaller planetary bodies, including asteroids. Models of the early Solar System (ESS) (e. g., Safronov, 1969; Lewis, 1974; Makalkin, Dorofeeva, 2009; Bitsch et al., 2015) as well as taxonomic classifications of asteroids based on their spectral observations (Tholen, 1984; Bell et al., 1989; Bus, Binzel, 2002a, b) show that the snow-line at the stage of planetesimal formation was somewhere between the outer edge of the Main asteroid belt (or MAB, ~3.3 AU) and Jupiter's orbit (5.2 AU). However, the boundary could not stabilize in the ESS because of probable luminosity changes in the young Sun and decreasing turbulent viscous dissipation (e. g., Baraffe et al., 2015; Bitsch et al., 2015). Besides, due to intense ejection of residual stone-ice bodies by Jupiter from its formation zone to the inner region (Safronov, 1969; Safronov, Ziglina, 1991), it might has been possible to deliver icy material to the MAB (Busarev, 2002, 2011, 2012, 2016). For the two last plausible factors, it could result in mixing substances of silicate and ice composition in the MAB. Another important process, as follows from investigations of chondritic meteorites, was the early thermal evolution of silicate planetary bodies due to decay of short-lived isotopes (predominately $^{26}Al$ with $T_{1/2}= 0.72$ My) (e. g., Grimm, McSween, 1993; Srinivasan et al., 1999). According to our analytical estimations (Busarev et al., 2003), for the first few million years after formation of calcium-aluminum inclusions (CAIs) in the silicate fraction of matter, internal water oceans (at an average temperature of ~4 °C) could form into big enough stone-ice bodies ($R >$100 km) everywhere beyond the snow-line including the formation zone of proto-Jupiter. The main consequences of the processes would be aqueous differentiation of such bodies and accumulation of a silicate-organic core (up to ~0.7 $R$) saturated with liquid water and heavy organics (of kerogen or bitumen type with density > 1 g/cm$^3$) (Busarev et al., 2003, 2005). According to numeric modeling, important related processes in the water-rock system of the bodies were likely the exothermic reaction of serpentinization (hydration) of silicates and intense release of $H_2$ and $CH_4$ gases (Wilson et al., 1999; Rosenberg et al., 2001). These processes resulted in layered, microporous, and fragile internal structures of the bodies, and therefore, their low mechanical strength. It could be one of the main reasons for the predominant crushing rock-ice bodies from Jupiter's zone in collisions with asteroid parent bodies. Thus, all the above processes could lead to formation of a broad heliocentric distribution of the most abundant primitive type asteroids (C-type and similar) with a maximum near ~3 AU (Bell et al., 1989; Bus, Binzel, 2002b; Busarev, 2011, 2012; Alexander et al., 2012). It suggests that most, if not all, main-belt primitive asteroids include a considerable icy component.

Previous findings of cometary-like bodies among main-belt asteroids were interpreted in most cases as random events connected with "dynamical" contamination of the asteroid family with atypical icy objects (for instance, extinct comet nucleus) which become active only due to sporadic collisions or impacts (e. g., Hsieh, Jewitt, 2006; Hsieh, Haghighipour, 2016). Another point of view based on modeling and observations is that free water ice is widespread originally in the subsurface interiors of primitive main-belt asteroids themselves (e. g., Schorghofer, 2008; Rivkin, A. S., Emery, J. P., 2010 ). Our discovering simultaneous sublimation activity on several main-belt primitive asteroids at shortest heliocentric distances supports the last opinion and points likely to the same or similar physical and chemical conditions of origin of the bodies corresponding to the outer edge of the main-belt and beyond.

## 2. Observations, data reduction and asteroid reflectance calculation

Observational data on the asteroids were already described in more details in our previous publication (Busarev et al., 2015b) along with a comparison with previously obtained data by other authors. It is discussed here only briefly. Spectra of the main-belt asteroids (145) Adeona, (704) Interamnia, (779) Nina, and (1474) Beira were obtained in September 2012 using a 2-m telescope with a low-resolution (R ≈ 100) CCD-spectrophotometer in the wavelength range 0.35-0.90 μm at Terskol Observatory (Mt. Terskol, 3150 m above sea level, Russia) operated by IA RAS. Ephemerides and some observational parameters of Adeona, Interamnia, Nina, and Beira (calculated with on-line service of IAU Minor Planet Center http://www.minorplanetcenter.net/iau/MPEph/MPEph.html) are given in Table 1 at the times (UT) corresponding to averaged reflectance spectra of the asteroids on each night (Figs 1-5). DECH spectral package (Galazutdinov, 1992) was used to reduce CCD observations by means of standard reduction procedures (flat-field correction, bias and dark subtraction, etc.) and to extract asteroid spectra. Wavelength calibration of the spectra was done using the positions of hydrogen Balmer lines in the spectrum of α Peg (B9III) observed in a repeated mode. The obtained spectra were corrected for the difference in air mass by applying a conventional method based on observations of a standard star being, as well, a solar analog (e.g., Hardorp, 1980). Its spectrum was also taken (instead of that for the Sun) for calculation of asteroid reflectance spectra (e.g., McCord et al., 1970). Observational data on the same star were used to determine the running spectral extinction function of the terrestrial atmosphere. Altogether two reflectance spectra of Adeona, five of Interamnia, twelve of Nina (on 13 September 2012), and five (on 18 September 2012) and nine (on 19 September 2012) of Beira were obtained and averaged on corresponding nights and shown in Figs 1-3 and 5. Moreover, we here compare reflectance spectra of (779) Nina obtained on 13 September 2012 with new spectra obtained on 26-28 September 2016 near perihelion distance of the asteroid (Fig. 4) to confirm its repeated sublimation activity. It is important to note that Nina's reflectance spectra on 26 and 28 September 2016 correspond to nearly the same rotational phases of the asteroid at its decreasing heliocentric distance. We interpret the short-wavelength differences in the shape of reflectance spectra of Nina obtained in 2012 and 2016 (Fig. 4) as a result of different intensity of sublimation activity of the asteroid before and after perihelion. But this in more detail below.

### 3. Interpretation of obtained asteroid reflectance spectra

### 3.1. (145) Adeona, (704) Interamnia, (779) Nina, and (1474) Beira

Some important characteristics of the asteroids are as follows. An average diameter of Adeona is estimated in the range from 126.13 km (Masiero et al., 2014) to 151.14 km (Tedesko et al., 2004), rotational period of 15.071 h (Harris et al., 2012) and orbital period of 4.37 yr (http://ssd.jpl.nasa.gov/sbdb.cgi#top). An average diameter of Interamnia is estimated in the range from 307.31 km asteroid (Masiero et al., 2014) to 316.62 km (Tedesko et al., 2004), rotational period of 8.727 h (Harris et al., 2012) and orbiting the Sun with a period of 5.35 yr (http://ssd.jpl.nasa.gov/sbdb.cgi#top). Nina's average diameter is estimated in the range from 76.62 km (Tedesko et al., 2004) to 80.57 km (Masiero et al., 2014). It rotates with a period of 11.186 h (Harris et al., 2012) and orbits the Sun for 4.35 yr (http://ssd.jpl.nasa.gov/sbdb.cgi#top). Finally, Mars crosser Beira has rotational period of 4.184 h and orbits the Sun for 4.52 yr (http://ssd.jpl.nasa.gov/sbdb.cgi#top). An average diameter of Beira is estimated as ~10 km (Busarev et al., 2016).

Table 1. Ephemerides and some observational parameters of (145) Adeona, (704) Interamnia, (779)

| 145 Adeona | | | | | | | | | | |
|---|---|---|---|---|---|---|---|---|---|---|
| Date (y m d) | UT middle ( hms ) | R.A. (h m) | Decl. (° ´ ) | Delta (AU) | r (AU) | Elong.(°) | Ph.(°) | V(m) | Elev.(°) | Exp. Time (s) | Air Mass (M) |
| 2012 09 19 | 231000 | 02 29.58 | -01 38.1 | 1.843 | 2.692 | 140.1 | 13.9 | 12.4 | 44.0 | 1200 | 1.4383 |
| Solar analog star HD10307 (R.A. = 01 41.78;  Decl. = +42 36.8): Elev. 67´ at t=01ʰ 32ᵐ ,  M = 1.0862 | | | | | | | | | | |

| 704 Interamnia | | | | | | | | | | |
|---|---|---|---|---|---|---|---|---|---|---|
| Date (y m d) | UT middle ( hms ) | R.A. (h m) | Decl. (° ´ ) | Delta (AU) | r (AU) | Elong.(°) | Ph.(°) | V(m) | Elev.(°) | Exp. Time (s) | Air Mass |
| 2012 09 13 | 213300 | 03 24.64 | +39 00.3 | 2.080 | 2.616 | 111.1 | 21.0 | 10.9 | 51 | 600 | 1.2896 |
| Solar analog star HD10307 (R.A. = 01 41.78;  Decl. = +42 36.8): Elev. 48´ at t=19ʰ 25ᵐ, M = 1.3446 | | | | | | | | | | |

| 779 Nina | | | | | | | | | | |
|---|---|---|---|---|---|---|---|---|---|---|
| Date (y m d) | UT middle (hms ) | R.A. (h m) | Decl. (° ´ ) | Delta (AU) | r (AU) | Elong.(°) | Ph.(°) | V(m) | Elev.(°) | Exp. Time (s) | Air Mass |
| 2012 09 13 | 195020 | 00 53.05 | +32 41.0 | 1.289 | 2.149 | 138.6 | 18.0 | 11.0 | 55.2 | 300 | 1.3201 |
| Solar analog star HD10307: Elev. 48´ at t=19ʰ 25ᵐ, M = 1.3446 | | | | | | | | | | |
| 2016 09 26 | 231000 | 20 00.42 | -05 21.4 | 1.413 | 2.060 | 116.1 | 25.9 | 11.4 | 18.6 | 1500 | 2.8688 |
| Solar analog star HD9986   (R.A. = 01 37 41;  Decl. = +12 04 42): Elev 24´  at t=22ʰ 50ᵐ, M = 2.447 | | | | | | | | | | |
| 2016 09 28 | 230000 | 20 01.72 | -05 17.85 | 1.431 | 2.060 | 114.5 | 26.3 | 11.4 | 23.3 | 400 | 2.5347 |
| Solar analog star HD173071   (R.A. = 18 43 15.8;  Decl. = +09 02 28.6): Elev 24´ at t=22ʰ 50ᵐ, M = 2.447 | | | | | | | | | | |

| 1474 Beira | | | | | | | | | | |
|---|---|---|---|---|---|---|---|---|---|---|
| Date (y m d) | UT middle (hms ) | R.A. (h m) | Decl. (° ´ ) | Delta (AU) | r (AU) | Elong.(°) | Ph.(°) | V(m) | Elev.(°) | Exp. Time (s) | Air Mass |
| 2012 09 18 | 205300 | 22 32.88 | +32 60.4 | 0.651 | 1.567 | 141.4 | 23.6 | 13.8 | 73.3 | 1200 | 1.047 |
| Solar analog star HD10307 (R.A. = 01 41.78;  Decl. = +42 36.8): Elev. 63´ at t=01ʰ 30ᵐ,  Decl. = 1.122 | | | | | | | | | | |
| 2012 09 19 | 190413 | 22 30.72 | +33 24.3 | 0.649 | 1.563 | 140.8 | 24.0 | 13.8 | 72.7 | 1200 | 1.055 |
| Solar analog star HD10307 (R.A. = 01 41.78;  Decl. = +42 36.8): Elev. 61´ at t=01ʰ 30ᵐ,  Decl. = 1.143 | | | | | | | | | | |

Nina, and (1474) Beira.

Description of designations: UT – universal time; R.A. – right ascension; Decl. – declination; Delta – distance from the Earth's center to the object's center; r – distance from the object's center to the Sun's center; Elong. – elongation of the object; Ph. –

According to popular classifications (Tholen, 1989; Bus and Binzel, 2002b), Adeona is of C or Ch type, Interamnia – of F or B, Nina – of M or X, and Beira – of FX or B. The values of the geometric V-band albedo of Adeona, Interamnia, and Nina are 0.06, 0.08, and 0.16, respectively (Masiero et al., 2014). Taking into account the B classification of Beira (Bus, Binzel, 2002b), we conditionally adopted its geometric albedo (still unknown) as ~0.08. It is important to note that such type asteroids are supposed to be primitive, including low-temperature compounds (hydrated silicates, oxides and hydro-oxides, organics, etc.) (e. g., Gaffey et al., 1989, 2002). Despite the previously debated high-temperature mineralogy of Nina, radar observations showed that the asteroid is also primitive but has a heterogeneous composition (Shepard et al., 2010).

We discovered (Busarev et al., 2015a, b) unusual overall shapes of the reflectance spectra of Adeona, Interamnia, Nina (in the range ~0.4-0.6 μm), and Beira (in the range ~0.55-0.75 μm) at the times of observations manifesting itself in a considerable growth of reflectivity in the short-wavelength range (up to ~30-40%) (Figs 1-3, 5). At the same time, such a maximum was absent in reflectance spectra of other asteroids observed by us at the Terskol observatory with the same facilities and nearly concurrently. More subtle spectral features of Adeona, Interamnia, Nina, and Beira corresponding to their surface mineralogy (located approximately at 0.39, 0.46 and 0.71 μm) were analyzed, as well (Busarev et al., 2015a, 2015b). Among them, Nina demonstrated considerable spectral differences with rotation. Spectral heterogeneity of Nina's surface material is confirmed by elevated standard deviations in the range of 0.55−0.92 μm as seen in its average reflectance spectrum obtained (from 13 separate reflectance spectra) over about a half of the rotational period (Fig. 3).

It is important to emphasize that in time of our observations, Interamnia, Nina, and Beira were near their perihelion distances, when temperature on the surface reached a maximum near the asteroids' subsolar points. At the same time, Adeona was approaching the Sun and close to its middle heliocentric distance. It was suggested that the unusual visible-range rise of asteroid reflectivity could be a result of sublimation activity of the surface matter including ices (Busarev et al., 2015a). A subtle cloud surrounding the asteroid, or a coma of sublimed and frozen particles, could produce scattering of the light reflected from the asteroid surface. As seen in Figs 1-3 and 5, this rise takes place around ~0.4-0.6 μm or ~0.55-0.75 μm. In addition, our analysis of these reflectance spectra showed that scattering of reflected light in the coma does not change position of intrinsic absorption bands produced by asteroid mineralogy (Busarev et al., 2015a).

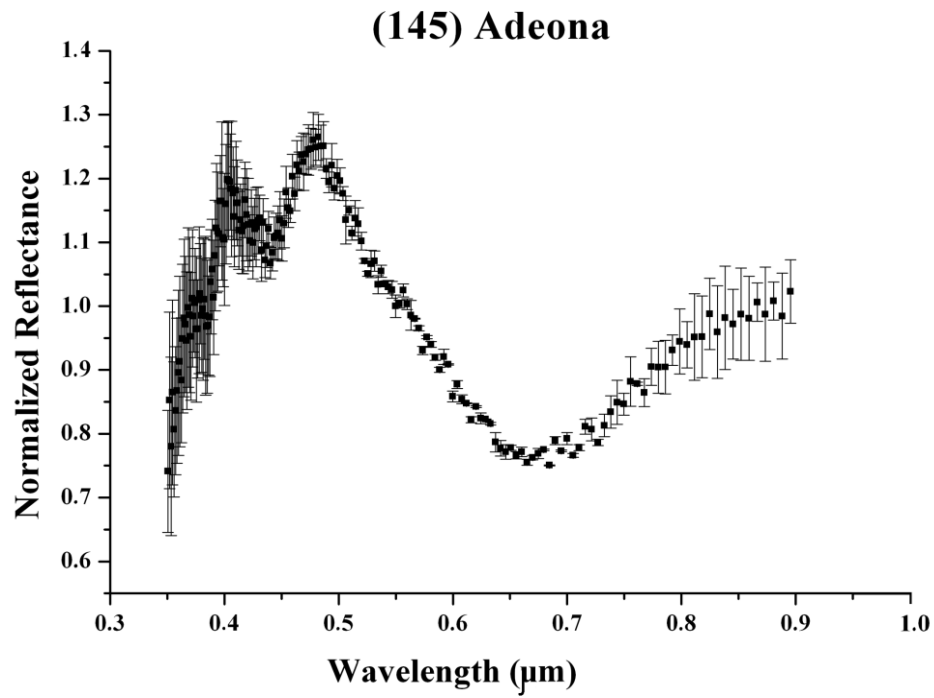

Fig. 1. Averaged and normalized ($R = 1.0$ at 0.55 µm) reflectance spectrum of (145) Adeona on 19 September 2012. Error bars represent the standard deviation.

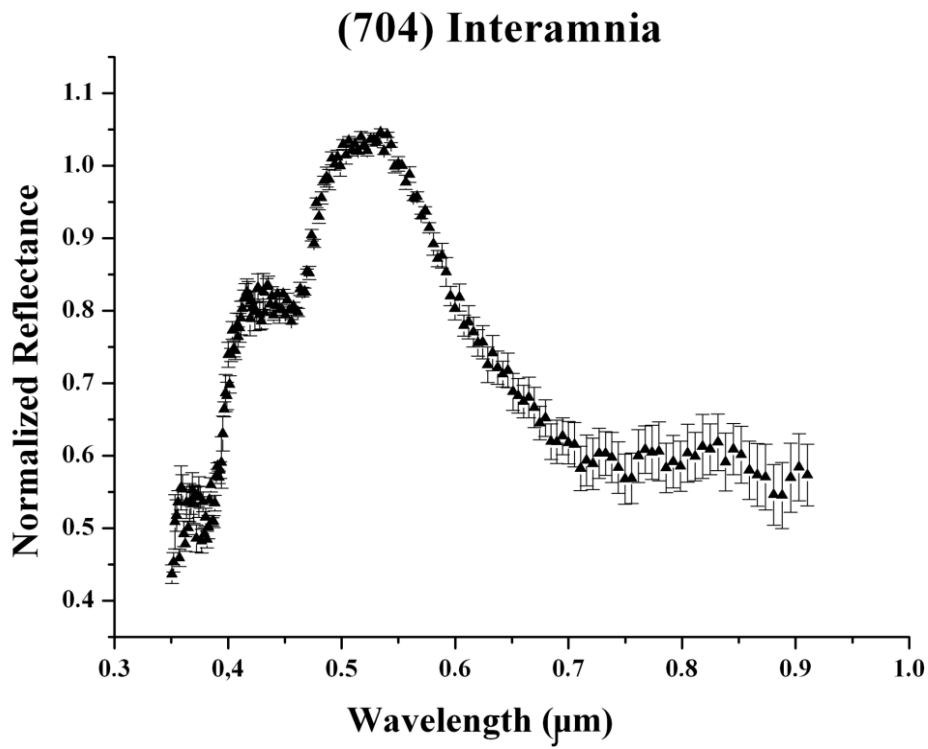

Fig. 2. Averaged and normalized ($R = 1.0$ at 0.55 µm) reflectance spectrum of (704) Interamnia on 13 September 2012. Error bars represent the standard deviation.

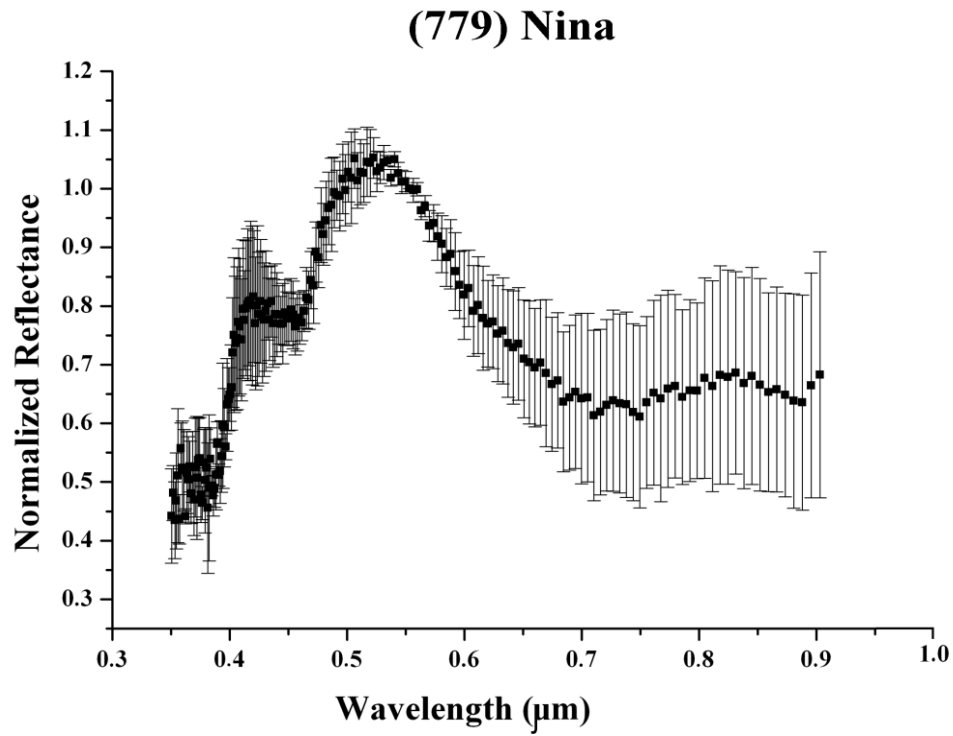

Fig. 3. Averaged and normalized ($R$ = 1.0 at 0.55 µm) reflectance spectrum of (779) Nina on 13 September 2012. Error bars represent the standard deviation.

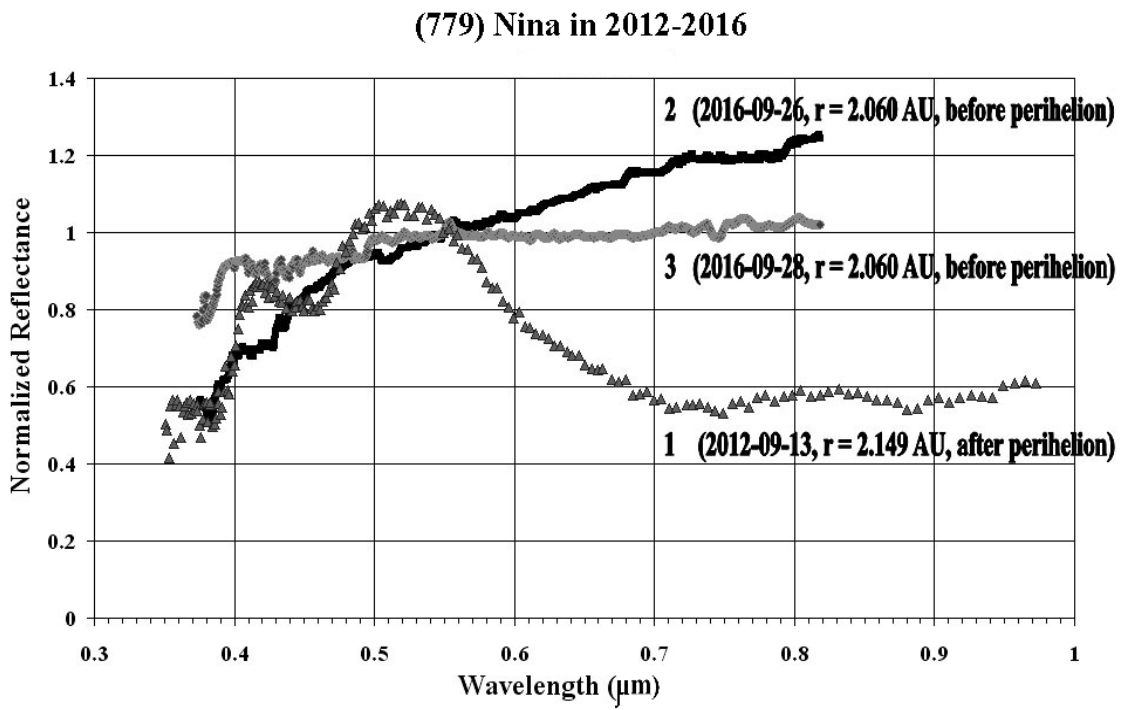

Fig. 4. Comparison of averaged and normalized ($R$ = 1.0 at 0.55 µm) reflectance spectra of (779) Nina obtained on 13 September 2012 and 26-28 September 2016 near perihelion distance of the asteroid. Spectra on 26 and 28 September 2016 were obtained nearly at the same rotational phases of Nina and correspond to its successively decreasing heliocentric distance near perihelion (q = 2.0589 AU).

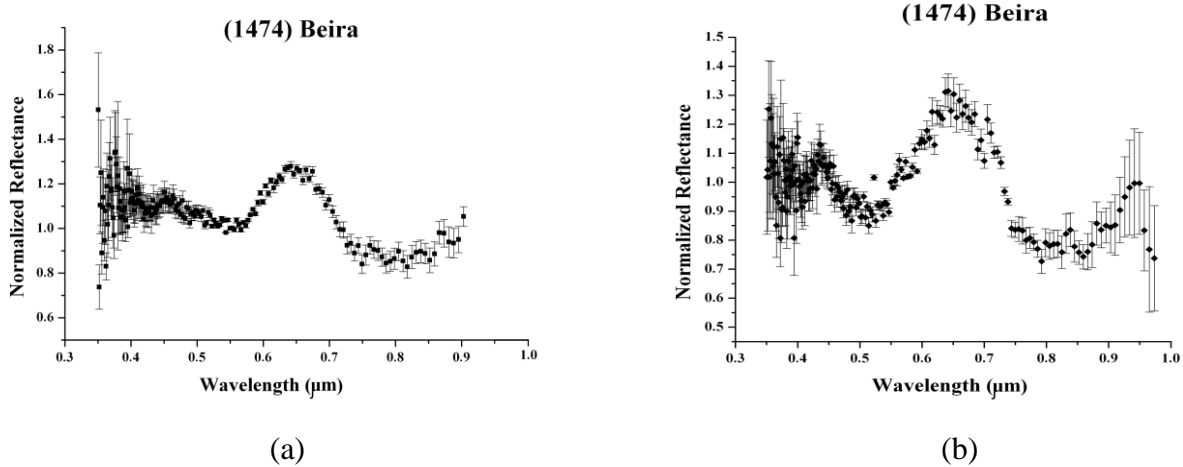

(a)                                                                        (b)

Fig. 5 (a, b). Averaged reflectance spectra of (1474) Beira normalized at 0.55 µm: (a) – on September 18 2012, (b) – on 2012 September 19. Error bars represent the standard deviation.

Later, on 26 and 28 September 2016, we acquired spectral confirmations of sublimation activity of (779) Nina during its last approach to perihelion. Nina's reflectance spectra obtained at gradually decreasing heliocentric distances (Table 1) demonstrate a dramatic change of the spectral slope from positive to neutral (nearly at the same rotational phases) produced likely by a growth of coma of sublimed ice particles scattering reflected from the asteroid light in the short-wavelength range (Fig. 4, curves 2 and 3).

Additional indirect confirmations of likelihood of the reflectance spectra of Adeona, Interamnia, and Nina we consider obtained with *Dawn* spacecraft (NASA) reflectance spectra of (1) Ceres' plots with subliming ice material (or bright spots) at the bottom of some large craters on (Nathues et al., 2015). There is obvious similarity of all these spectra (Figs 1-3 and Fig. 6, upper three curves). The authors (Nathues et al., 2015) relate the unusual shape of the Ceres' reflectance spectra with maximum at 0.55 µm produced by scattering reflected from the surface sunlight in a cloud of sublimed $H_2O$ ice particles. Indeed, a weak coma of water vapor around Ceres at perihelion was discovered previously in the UV range by the *International Ultraviolet Explorer satellite* (NASA, UK SRC, ESA) (A'Hearn, Feldman, 1992) and in the far IR range by the *Herschel Space Observatory satellite* (ESA) (Küppers et al., 2014). A flow rate of water vapor molecules from the entire surface of Ceres was estimated as $10^{26}$ particles/s (Küppers et al., 2014). To explain Ceres' $H_2O$ plumes, several mechanisms have been discussed, and a study of dynamic evolution of the plumes led to a conclusion that cometary-like sublimation is preferable because of a correlation between the magnitude of the emission and the change in the heliocentric distance along the orbit (Formisano et al., 2016).

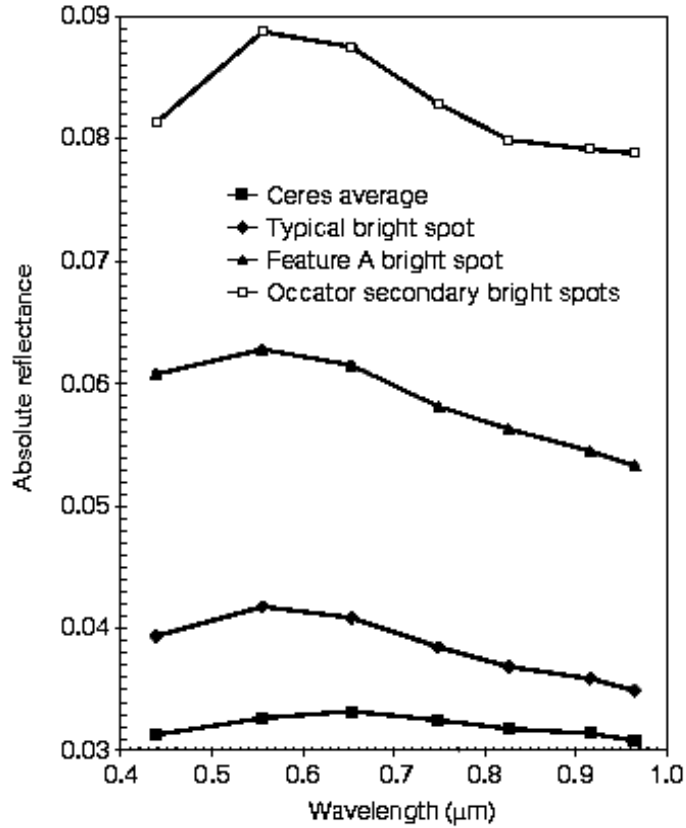

Fig. 6. Reflectance spectra of (1) Ceres average (C-type) and its bright spots at the bottom of some large craters obtained by *Dawn* spacecraft (NASA). The figure is reproduced from Nathues et al. (2015) with permission of authors.

### 3.2. Near-Earth asteroid (162173) Ryugu

To our surprise, we found out a similarity of the reflectance spectra of four discussed main-belt asteroids in time of their sublimation activity and one of (162173) Ryugu, a potentially hazardous near-Earth asteroid (NEA) of Apollo's family, observed by F. Vilas in 2007 (Vilas, 2008). Main parameters of the asteroid are as follows: diameter and geometric albedo ~0.87 km and ~0.07, respectively (Ishiguro et al., 2014), classified as Cg-type asteroid (Binzel et al., 2001, 2002), and its orbital and rotation periods are 1.3 yr and 7.627$^h$ (http://ssd.jpl.nasa.gov/sbdb.cgi#top). As dynamical modeling shows, the asteroid could be delivered to near-Earth space from the inner main-belt region (2.15 AU < $a$ < 2.5 AU, $i$ < 8º) due to the action of $v_6$ resonance (Campins et al., 2013). Ryugu had been selected as the target of Japan's *Hayabusa 2* space mission providing a sample return of the asteroid in 2020 (Ishiguro et al., 2014). Spectral observations of Ryugu have been performed over three nights, 2007 July 11, September 10, and September 11, at the MMT Observatory (Mt. Hopkins, 2617-m elevation, USA) 6.5 m telescope and the facility Red Channel CCD spectrograph in the range 0.50-0.90 µm, and the reflectance spectra are given in Fig. 7 (Vilas, 2008). The reflectance spectrum on 11 July 2007 obtained nearly a month after Ryugu's passage of aphelion has unusual shape and a maximum at ~0.57 µm making it very similar to those of considered asteroids (Fig. 7, top spectrum). Obviously, another important feature of the spectrum is a considerable dispersion of its points over the full spectral range, which could be produced by a scattering medium between the asteroid and observer.

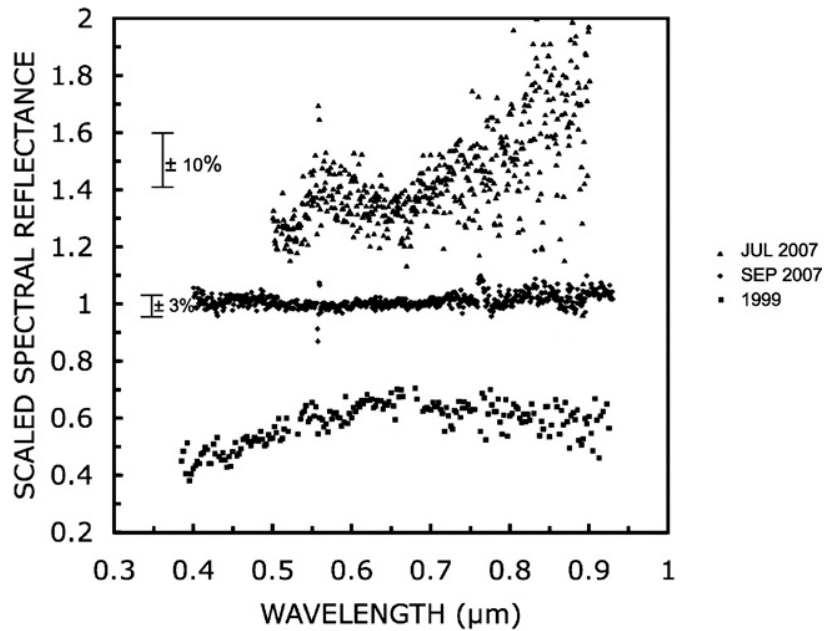

Fig. 7. Normalized ($R$ = 1.0 at 0.55 µm) and offset reflectance spectra of (162173) Ryugu (its values of perihelion $q$ = 0.963292 AU and aphelion $Q$ =1.415819 AU) reproduced from article of Vilas (2008). Two upper spectra were obtained on 2007 July 11 (heliocentric distance $r$ = 1.376 AU, after aphelion) and 2007 September 10/11 (composite, $r$ = 1.241 AU, after aphelion), respectively. Error bars represent scattering of points in the spectra. The lowermost spectrum of Ryugu was obtained during its 1999 discovery apparition by Binzel et al. (2001) (17/05/99, $r$ = 1.311 AU, before aphelion).

A striking difference of Ryugu's reflectance spectrum on July 11 from that on September 10/11 2007 was previously interpreted as a result of large variegation of asteroid surface mineralogy with rotation and, in particular, presence of CM-like 0.7-micron band sometimes (Vilas, 2008). However, new reflectance spectra of Ryugu obtained in a wide range of rotational phases (Lazzaro et al., 2013; Sugita et al., 2013) and their comparison with reflectance spectra of Murchison meteorite (CM2-group) heated at several temperatures in the range 550-900 ℃ (Sugita et al., 2013; Moskovitz et al., 2013) did not confirm presence of strong mineralogical variations or irregular temperature metamorphism on the asteroid which could have spectral manifestations comparable in magnitude with that found by Vilas (Vilas, 2008).

At the same time, spectroscopic observations of Ryugu on 5 July 2012, just before its aphelion passage, showed an elevated spectral reflectance of Ryugu in the short-wavelength range (Sugita et al., 2013, Fig. 1 from the mentioned publication).

**4. Interpretation of unusual reflectance spectra of (145) Adeona, (704) Interamnia, (779) Nina, (1474) Beiram, and (162173) Ryugu**

Verifying our assumption about sublimation activity on/under the surface of observed primitive asteroids, we estimate changes in the subsolar temperature. Assuming they emit thermal radiation nearly as black bodies (e. g., Harris and Lagerros, 2002), the surface temperature is described by the Stefan-Boltzmann law. Additionally, we relate the received solar electro-magnetic energy by the unit area at an arbitrary top level of the terrestrial atmosphere at the subsolar point (or the solar constant) and the radiative energy received by the unit area normal to the direction to the Sun at heliocentric distance $r$. Taking into account the values of Stefan-Boltzmann's and solar constants, we obtained a simple formula for estimation of instantaneous effective temperature in the subsolar point ($T_{ss}$) on equator of an atmosphereless solid body with zero thermal inertia located at heliocentric distance $r$ (in AU) (Busarev et al., 2015a):

$$T_{ss} = 394 \text{ K} \cdot ((1 - p_v) / r^2)^{1/4},$$ (1)

where $p_v$ is the geometric albedo.

Calculations using this formula give changes of the subsolar temperature on the asteroids with time (and heliocentric distance) including moments of the observations (Figs 8-11). These specific values are next: $T_{ss, o}$ = 236.5 K (Adeona), $T_{ss, o}$ = 238.6 K (Interamnia), $T_{ss, o}$ = 257.3 K (Nina), and $T_{ss, o}$ = 308.3 K (Beira). As seen, Adeona's and Interamnia's values are very close. The temperature on Adeona, Interamnia, and Nina at time of observations is in the range of the most intense sublimation of water ice, namely, within several tens of degrees below 273 K. While that on Beira exceeded the melting point of water ice by 35 degrees.

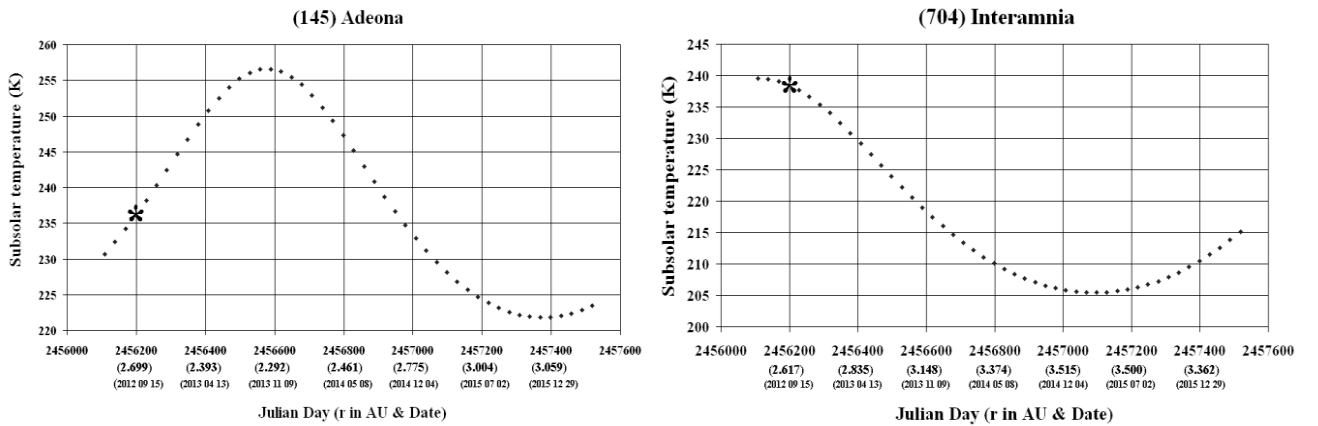

Fig. 8. Subsolar (equatorial) temperature changes on the surface of (145) Adeona while orbiting the Sun. Temperature and heliocentric distance of the asteroid at the time of observations (Julian Day) are marked by asterisk.

Fig. 9. Subsolar (equatorial) temperature variations on the surface of (704) Interamnia while orbiting the Sun. Temperature and heliocentric distance of the asteroid at the time of observations (Julian Day) are marked by asterisk.

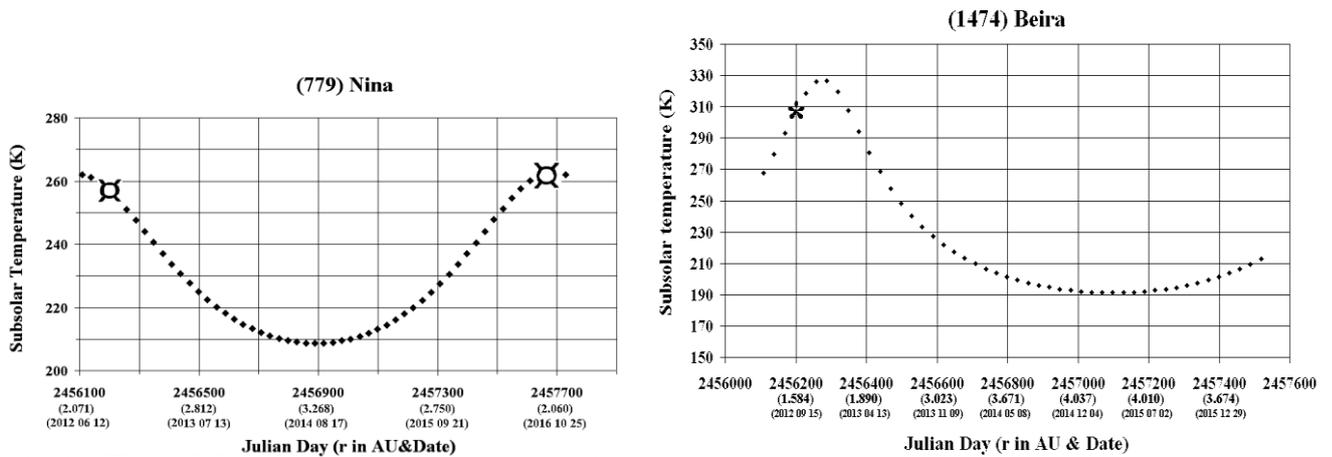

Fig. 10. Subsolar (equatorial) temperature variations on the surface of (779) Nina while orbiting the Sun. The temperature and heliocentric distance of the asteroid at the time of observations (Julian Day) are marked by open crosses.

Fig. 11. Subsolar (equatorial) temperature variations on the surface of (1474) Beira while orbiting the Sun. The temperature and heliocentric distance of the asteroid at the time of observations (Julian Day) are marked by asterisk.

As follows from the calculations, the ranges of subsolar temperature are: ~222÷257 K (on Adeona), ~206÷240 K (on Interamnia), ~208÷262 K (on Nina), and ~191÷329 K (on Beira) (Figs 8-11). For comparison, the calculated subsolar temperature on the surface of Ryugu is within ~326÷394 K (Fig. 12) exceeding at the maximum the temperature of boiling water by twenty degrees.

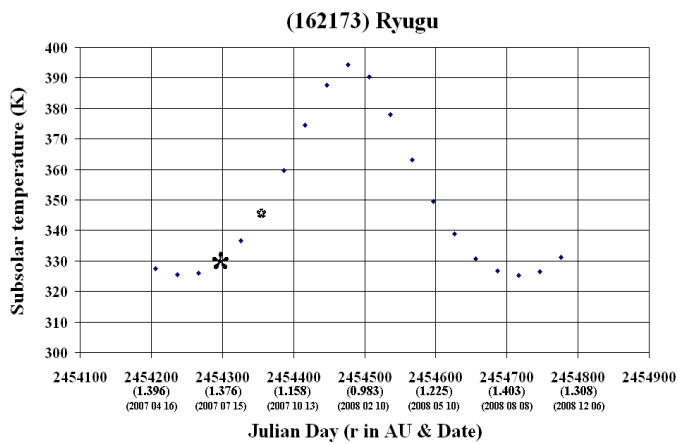

Fig. 12. The subsolar temperature on (162173) Ryugu with orbiting the Sun. Values of temperature and heliocentric distance of the asteroid at the time of observations (Vilas, 2008) are marked by asterisks: large one on 2007 July 11, the instant of possible sublimation activity; and small one on 2007 September 10/11.

Among volatiles most abundant near planetary surfaces (e. g., Dodson-Robison et al., 2009), the triple point in the phase diagram for $H_2O$ and $CO_2$ (and their mixtures) is at the closest position to this temperature ranges. However, the point of $CO_2$ is shifted downward by ~50 degrees (e. g., Longhi, 2005). Therefore, one could expect that $H_2O$ predominantly survives under the surface of primitive asteroids since their formation or since previous falls of smaller ice bodies on the asteroids (e. g., Fanale and Salvail, 1989; Jewitt and Guilbert-Lepoutre, 2012).

Even a subtle cloud or coma of sublimed micron-sized particles surrounding an asteroid could produce a considerable scattering of light reflected from the asteroid. Given the spectrally neutral refractive index of the particles in the used range (as for water ice and "dry ice"), the wavelength position of a "hump" in asteroid reflectance spectrum is likely determined by a predominant particle size in the scattering coma. Actually, as follows from Mie theory and subsequent numeric modeling (Mie, 1908; Hansen and Travis, 1974), for particles with refractive index of water ice ($n = 1.33$), intensity of scattering is maximal for the dimensionless parameter $x = (2\pi \cdot \rho / \lambda) \approx 6$ or $\rho / \lambda \approx 1$, where $\rho$ is radius of scattering particles. An average radius of scattering particles ($\rho$) and spectral position of maximum of the scattered light ($\lambda$) are matching in this case. Then, from aforementioned theoretical results and our measurements, we could assess an average radius of ice particles in comas of considered asteroids as these values: ~0.48 μm (Adeona, Fig. 1), ~0.53 μm (Interamnia, Fig. 2), ~0.52 μm (Nina, Fig. 3), and ~0.65 μm (Beira, Fig. 5). Similarly, we estimate the value for Ryugu (Fig. 6, top spectrum) as ~0.57 μm. As seen from the values, submicron-sized particles might have been in average present only in Adeona's coma. Apparently, more high temperature conditions of ice sublimation (e. g., origin of more strong vapor streams) on Beira and Ryugu could lead to carrying out of larger particles (ice and dust or their conglomerates) to the asteroid coma.

However, we do not exclude possibility of another interpretation of the phenomenon. Given a predominately CM-like composition of primitive asteroids, the varying strength of absorption features at 0.7-micron instead of Mie scattering of fine particles could make a positive peak in asteroid reflectance near 0.5-0.6 μm. This may be especially relevant for NEAs like Ryugu with the subsolar temperature well above the boiling temperature of water near perihelion. Formation of the near-surface water vapor streams could considerably resurface material of such NEAs and, as a result, strengthen the intensity of CM-like 0.7-micron band hidden other times owing to space weathering or heating. Though, the mechanism could act along with Mie scattering in the asteroid coma.

## 5. Possible mechanisms of activity on the main-belt and near-Earth asteroids

Simultaneous sublimation activity of main-belt primitive asteroids Interamnia, Nina, and Beira near perihelion distances may be crucial. Though approaching the Sun, Adeona was near its mid heliocentric distance, and the subsolar temperature on its surface (and subsurface thermal conditions) was very close to that of Interamnia. Taking into account Adeona's taxonomic type and very low geometric albedo, its surface material could be the most primitive among the considered bodies, and could include a considerable proportion of ices. Thus, the rise of the subsolar temperature when the asteroids approach the Sun should be a main cause of their sublimation activity. The phenomenon could also be an indication of a sizeable ice component in C-type and similar asteroids. A synchronism in sublimation activity on several main-belt primitive asteroids with shortening heliocentric distances points to similar conditions of their origin and evolution. It is in agreement with the overall heliocentric distribution of C-type asteroids with a maximum shifted to external edge of the main asteroid belt (Bell et al., 1989; Bus, Binzel, 2002b) being closer to the assumed position of a snow-line in the ESS.

Due to early aqueous differentiation of volatile-abundant matter of primitive asteroids (or their parent bodies) because of decay of short-lived radionuclides ([26]Al, [60]Fe, etc.) (Srinivasan et al., 1999; Ghosh et al., 2006; Wadhwa et al., 2006), the bodies accumulated ices in the subsurface layers in the shape of an ice mantle (Grimm and McSween, 1989, 1993; Busarev et al., 2003, 2005). The found Adeona's, Interamnia's, and Nina's subsolar temperature ranges (Figs 8-10) are below the melting point of water ice but they are higher than the threshold temperature for $H_2O$ ice sublimation (~145 K) (Schorghofer, 2008). This means that primitive main-belt asteroids have predominantly slowly and gradually deplete of the ice stock in the process of sublimation at shorter heliocentric distances (e. g., Schorghofer, 2008). In this respect, bodies like Beira constitute an exception among such asteroids: because of large eccentricity of their orbit, they experience more strong heating by the solar radiation near perihelion (see Fig. 11) and, hence, faster exhaustion of the subsurface ices. Regolith on primitive main-belt asteroids had likely been formed in a usual way (due to repeated impacts of meteoroids) but should also comprise ice inclusions if some ice layers existed in their interiors.

On the other hand, main-belt asteroids should have an ancient surface in average. Its common features are likely a considerable thickness of the regolith layer and its very low heat conductivity, similar to those of lunar regolith layer, which lead to damping of diurnal thermal wave in the surface layer of only several centimeters (e. g., Wechsler et al., 1972; Langset and Keihm, 1975; Kuzmin and Zabalueva, 2003). Thus, we could imagine on a qualitative level the internal structure of primitive main-belt asteroids as a combination of phyllosilicate and ice layers (and/or inclusions) with a thick regolith envelope (Fig. 12). In any case, sublimation activity on a primitive asteroid becomes likely more intense (not only at perihelion) after excavation of fresh ice by recent meteoroid impact(s). Because of asteroid rotation, such sublimation should take place predominately on the sunlit surface.

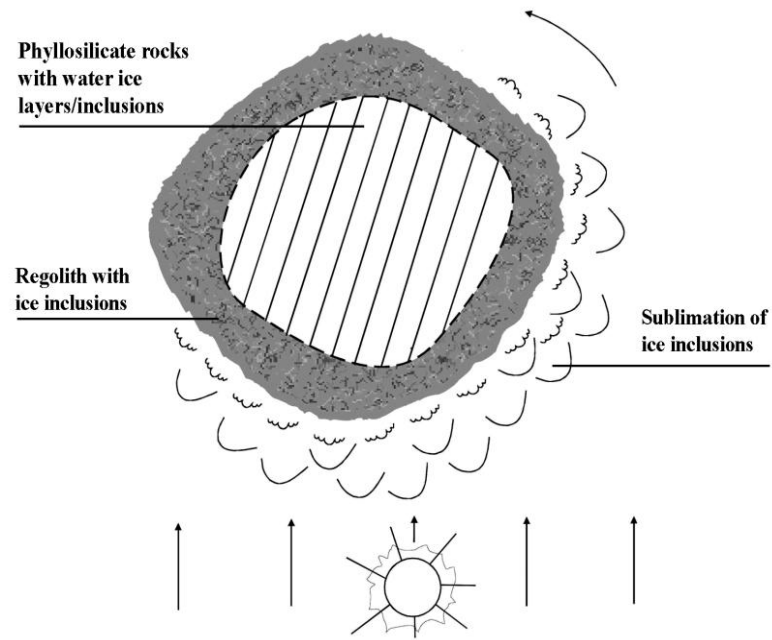

**Suggested internal structure
of a main-belt primitive asteroid**

Phyllosilicate rocks
with water ice
layers/inclusions

Regolith with
ice inclusions

Sublimation of
ice inclusions

Fig. 13. Suggested internal structure of a primitive main-belt asteroid and possible mechanism of its sublimation activity near perihelion.

Similar to the above reflectance spectra of the discussed main-belt asteroids, Ryugu's spectrum on 2007 July 11 (Vilas, 2008) has signs of scattering light reflected from the surface in the asteroid's coma formed by sublimed ice particles. However, the spectrum corresponds to the moment after passage of aphelion by the asteroid, when the subsolar temperature was close to a minimum (Fig. 12). It looks strange and contradicts the above scenario of sublimation activity of primitive main-belt asteroids. Nevertheless, one of Ryugu's reflectance spectra obtained five years later, on 5 July 2012 (Sugita et al., 2013), seems to confirm the onset of the process just before passage of aphelion by the asteroid.

Generally, it may be several sources of Ryugu's hypothetical activity. Possible sources of water ice could be in the Ryugu's interiors or cold traps on its surface. For implementation of the first possibility the ice should survive in the Ryugu's interiors for the period of the asteroids residence in the near-Earth orbit. When a main-belt primitive asteroid becomes a NEA due to a strong impact event or the action of gravitational resonances, it quickly (after several revolutions around the Sun) loses a considerable amount of ice stored near the surface due to considerable growth of its subsolar and average subsurface temperatures by ~100 degrees (e. g., Figs. 8−10 and 12). Tidal interactions with the Earth and other terrestrial planets could increase the number and sizes of fractures and breaks in the asteroid body that facilitate venting of any volatiles from its subsurface layers and loss of the uppermost regolith. A similar effect is produced probably by a sharp temperature difference between day and night sides of a NEA with its rotation on a timescale of several hours.

The upper subsurface layers of the asteroid are heated by solar radiation. In these layers the variable insolation generates temperature oscillations in the diurnal and seasonal skin layers. If we take ~350 K (an average subsolar temperature on Ryugu's surface over its orbital period taking into account a slower motion of the asteroid near aphelion than at perihelion) as an external maximum temperature (Fig. 12) and ~150 K as a minimum value of the surface temperature on its night side according to a thermophysical modeling for a similar object (Yu, Ji, 2015), this average internal temperature would be 240–250 K. This temperature is in fact the maximum temperature of the Ryugu's interiors. If Ryugu has arrived to the present orbit quite recently, it could preserve the mean temperature ~150 K, adopting an average thermal inertia about 200 J $K^{-1}m^{-2}s^{-1/2}$ (Müller et al., 2017), characteristic for internal water ice stocks of the main-belt C-type and similar asteroids (Schorghofer, 2016).

The characteristic timescale for heating the asteroid's interiors is

$$t_h \sim \frac{D^2}{12\kappa} = \frac{1}{12}\left(\frac{D\,\rho\,C}{\Gamma}\right)^2, \tag{2}$$

where $D \approx 870$ m is the diameter of Ryugu, $\kappa$ is its thermal diffusivity related to thermal inertia $\Gamma$, density $\rho$, and specific heat capacity $C$ by the equation $\Gamma = \rho C \sqrt{\kappa}$.

The estimated value of Ryugu's thermal inertia is in the range between 150 and 300 J $K^{-1}m^{-2}\,s^{-1/2}$ (Müller et al., 2017). The heat capacity is estimated as follows. As Ryugu is classified as a C type (Binzel et al., 2001, 2002), its material is thought to be chondritic and corresponds to the spectral characteristics to CM carbonaceous chondrites (Cloutis et al., 2011). For this case we adopt the widely used analytic fit of Yomogida and Matsui (1984) for heat capacity of the non-differentiated bulk chondritic material

$$C(T) = 800 + 0.25T - 1.5 \times 10^7\,T^{-2} \tag{3}$$

which is a good fit for the heat capacity of the CM2 chondrite (Cold Bokkerveld) (Opeil et al., 2010) as well as achondrite meteorite shergottite (Los Angeles) (Opeil et al., 2012).

Using previous numeric results of heat transport modeling in the material of similar bodies (Kuzmin, Zabalueva, 2003), we accept a relaxed value of temperature below the diurnal and seasonal skin layers as $T = 240$–250 K which is about the average subsurface temperature over Ryugu's orbital period. At $T = 240$ K Eq. (3) gives $C \approx 600$ J $kg^{-1}$ $K^{-1}$. The material density $\rho$ is assumed to be in the range from 1.3 g $cm^{-3}$ to 1.7 g $cm^{-3}$. The former value is average for C-type asteroids (Scheeres et al., 2015; Chesley et al., 2014), the latter is a bulk density of CM carbonaceous chondrites (e. g., Cold Bokkerveld) at an average porosity ~25% (Consolmagno et al., 2008; Opeil et al., 2010). The difference between two values is supposed to be due to the difference in macroporosity and water abundance of these two kinds of objects.

For the above values of the parameters involved in the Eq. (2), we obtain the heating timescale in the range $t_h = (1.4–9.2) \times 10^4$ yr. For the preferable (basic) set of parameters $\Gamma = 200$ J K$^{-1}$m$^{-2}$ s$^{-1/2}$, $\rho = 1.5$ g cm$^{-3}$, $C \approx 600$ J kg$^{-1}$ K$^{-1}$ we have $t_h \approx 4 \times 10^4$ yr. If Ryugu reached its current orbit more recently than the above time (measured from the present), it would retain significant mass of its primordial abundance of water ice.

However, if Ryugu arrived from the MAB earlier, its central region should be heated to the mean subsurface temperature $T = 240–250$ K. Nevertheless, even if Ryugu had enough time to lose most of its ice supplies, the ice in its central region still could survive, because this temperature is lower than the sublimation–condensation temperature at pressure existing in the centre of asteroid

$$P = (\pi / 6) G \rho^2 D^2,$$ (4)

where $G$ is the gravitational constant. This relation follows from the hydrostatic equilibrium equation.

For the pressure in the Ryugu's centre from Eq. (4) and the above range of density values we obtain $P = 45–76$ Pa. This value is far below that for the triple point of water, equal to 610 Pa, therefore liquid water cannot exist in the Ryugu's interiors. At vapor pressure of ice in the above range $P = 45–76$ Pa sublimation–condensation temperature of water ice is $T_w = 245–250$ K that is not lower than the asteroid's mean temperature $T = 240–250$ K. This relation allows certain amount of ice to remain in the Ryugu's centre if the permeability of the asteroid's material is sufficiently small.

If some amount of ice is preserved in the interiors of Ryugu, after sublimation of water vapor, it can anyway be delivered to the bottom of the subsurface layers as water vapor, possibly, by the use of forming microcracks. Less probable that the ice survived in these layers in the solid phase. A thermal mechanism of short-lived outbursts because of formation of deeper cracks and outgassing super volatile compounds was proposed in the course of the recent study of 67P/Churyumov-Gerasimenko comet (Skorov et al., 2016).

It can be supposed that because of the presence of the desiccated (ice-free) subsurface layer, the maximum of the subsolar temperature on the NEA at the moment of passage of perihelion still does not reach the water-containing (in the form of vapor or ice) layers. So one could expect a time delay (or "thermal lag") between the moment of maximum heating of the asteroid surface and that of the upper boundary of the water-containing reservoir.

The delay between perihelion passage by Ryugu and the onset of its suggested sublimation activity should include a time interval for the heat transport from the surface to the internal water vapor/ice reservoir ($t_1$) and a time for transfer of $H_2O$ vapor from the frozen interiors to the surface ($t_2$). Because of the relatively small size of the asteroid and its permeable structure of the upper (regolith) layer, we assume $t_1 \gg t_2$ and neglect the value of $t_2$. Taking into account known from spectroscopic data symmetrical character of Ryugu's possible sublimation process in time, a month before aphelion passage (Sugita et al., 2013) and nearly a month after (Vilas, 2008), we assume the delay of sublimation from perihelion passage is roughly equal to a half of its orbital period $P$ (Fig. 7, spectrum on 11 July 2007; Fig. 12, the moment of sublimation activity is marked by a large asterisk). As follows from these and other observations, after a relatively short time of ejection of water vapor, Ryugu becomes probably inactive again, up to its next passage of aphelion (Fig. 7, spectrum on 11/10 September 2007; Fig. 12, a small asterisk). Thus, we calculate the delay through the phase lag of the seasonal temperature oscillations. The phase lag forms when the temperature oscillations from the surface penetrate through the thermal skin layer. The amplitude of these oscillations $\Delta T$ decreases with depth $z$ exponentially (Turcotte, Schubert, 2002)

$$\Delta T(z) = \Delta T_s \exp(-z / d), \tag{5}$$

where $\Delta T_s$ is the amplitude of seasonal temperature oscillations on the asteroid's surface, $d$ is the thickness of the thermal skin layer, which for seasonal temperature oscillations is

$$d = \sqrt{\frac{P \kappa}{\pi}} = \frac{\Gamma}{\rho C} \sqrt{\frac{P}{\pi}} \, , \tag{6}$$

where $P$ is the orbital period which for Ryugu is equal 1.3 yr. Other parameters in Eq. (6) and their adopted values are explained above, after Eq. (2). The phase lag at depth $z$ is (Turcotte and Schubert, 2002) $\varphi = z / d$. In the case considered $\varphi = 2 \pi t / P = \pi$. This relation can be used to obtain the depth of the level which is reached by the thermal wave during a half of orbital period. It is equal to $z_1 = \pi d$, where the relation for $d$ is given by Eq. (6). The amplitude of temperature variations at this depth is given by Eq. (5) $\Delta T(\pi d) = \Delta T_s \exp(-\pi)$. For the case of Ryugu we obtain the thickness of the seasonal skin layer $d = 0.53$–$1.4$ m with the most probable value $d \approx 0.8$ m. For $z_1$ we have $z_1 = 1.7$–$4.4$ m with the most probable value $z_1 \approx 2.5$ m. The amplitude of seasonal temperature cycle $\Delta T_s \approx 60$ K gives $\Delta T(\pi d) \approx 3$ K at $z = z_1$ (the level which is reached by the thermal wave after the half of orbital period).

This result would mean that at the depth about 2–4 meters some reservoir of water vapor or ice exists (the latter is less probable). The reservoir is activated by the seasonal thermal wave with an amplitude of a few kelvins.

Thus, the presumed temporal sublimation/degassing activity of Ryugu points to the existence of a residual frozen core in its interior. It could be an indication of a relatively recent transition of the asteroid from the MAB to the near-Earth area. Ryugu's internal structure on a qualitative level could be shown as that in Fig. 14.

Another mechanism for sublimation on Ryugu may also be proposed. A similar observational pattern could origin in the case of residual ice presence near one of Ryugu's poles due to a considerable change of lighting conditions. According to recent estimations of spin-vector orientation derived from a combined analysis of data of visual lightcurves and mid-infrared photometry and spectroscopy (Müller et al., 2017), Ryugu has a retrograde rotation with the most likely axis orientation of $(\lambda, \beta)_{ecl} = (310°\text{-}340°, -40° \pm \sim 15°)$, a rotation period of $P_{sid} = 7.63109^{h}$, and a very low surface roughness (r. m. s. of surface slopes $< 0.1$). Then, one could imagine that at the mentioned asteroid parameters some ice-content material preserved in a surface depression (a cold trap) near the North Pole of the asteroid would be more sunlit near aphelion then at perihelion. Additional heating the icy material could produce observed sublimation activity of Ryugu in the vicinity of aphelion. However, owing to a relatively quick change of a NEA's orbital parameters under action of gravitational perturbations from terrestrial planets the latter mechanism could represent a temporal phenomenon providing that the residence time of the asteroid in the near-Earth space is not very long to keep the icy matter at the surface. This is so especially when taking into account a significant obliquity of Ryugu's rotational axis (~60°) (Müller et al., 2017). Besides, as thermogravimetric measurements (Frost et al., 2000) show, phyllosilicates (being the main constituents of matrix of carbonaceous chondrites) have three stages of dehydration with elevation of the temperature. Such compounds as ferruginous smectite and some nontronites lose a significant proportion of the mass at 300–360 K (12–17%), at 360–380 K (3–4%), and at 380–430 K (1–2%) (Frost et al., 2000). The first two temperature intervals are well correspond to the calculated range of subsolar temperatures of Ryugu ~326÷394 K (Fig. 12). So the first two stages of dehydration process in phyllosilicates may be an additional source of Ryugu's surface/subsurface activity.

Nevertheless, both the above mechanisms (depletion of internal water ice and sublimation of near-polar ice-deposits and dehydration of phyllosilicates) are possible and could give inputs to a sublimation/outgassing activity of Ryugu near aphelion.

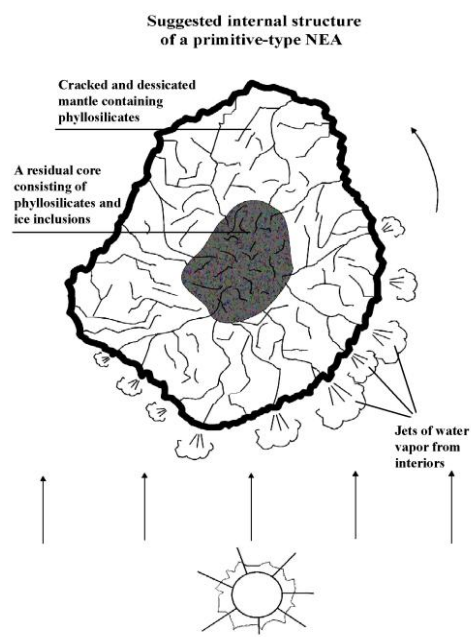

Fig. 14. Suggested internal structure of a primitive-type NEA including an icy core and possible mechanism of its sublimation/ degassing activity near aphelion.

## 6. Conclusions

The most interesting results of the work are the observed spectral signs of close in time similar sublimation activity of (145) Adeona, (704) Interamnia, (779) Nina, and (1474) Beira at shortest heliocentric distances. Taking into account their primitive nature, we propose that ice sublimation may be a widespread phenomenon among C and similar asteroids. So far we are able to confirm the periodicity of this phenomenon on the asteroid Nina (September 2016, Fig. 4). Confirmations should be made for the other discussed asteroids in the near future.

As is known, the mineralogy of solid rock bodies is sensitive to conditions of their origin. The same can be said about the content of volatile compounds in the matter. The orbital ellipticity of a small planet makes it possible to test the volatility of the body's material depending on heliocentric distance and accompanying surface temperature changes. The above results show that any main-belt asteroids that accumulated ices beneath the surface over their evolution could demonstrate such activity. Anyway, some transient sublimation activity on a primitive asteroid becomes more intense and/or prolonged in the case of excavation of buried $H_2O$ ice at recent meteoroid impact(s).

Suggested sublimation/degassing activity of (162173) Ryugu and possible comparatively small depth of its putative ice reservoir may indicate a relatively short residence time of the asteroid in the near-Earth space. This is particularly interesting in view of plans to visit the vicinity of Ryugu by *Hayabusa 2* spacecraft and delivery with its help of the asteroid sample on Earth.

## Acknowledgments

The observations of asteroids at the Terskol Observatory were supported by the Russian Foundation for Basic Research and Russian Science Foundation (projects No. 12-02-904 4 4-Ukr and 16-12-0 0 071, respectively). This study has been partially supported by the President of the RF grant NSh-9951.2016.2.

Authors of the paper thank anonymous reviewers for useful comments.